\documentclass[aps,prx,twocolumn,superscriptaddress,noshowpacs]{revtex4-2}

\usepackage{amsmath,amssymb}
\usepackage{graphicx}
\usepackage{dcolumn}
\usepackage{bm}
\usepackage{float}
\usepackage{placeins}
\usepackage{nicefrac,xfrac}
\usepackage{braket}
\usepackage{physics}
\usepackage{scalerel}

\usepackage[dvipsnames]{xcolor}
\usepackage{soul,color}
\usepackage[normalem]{ulem}
\usepackage{comment}
\usepackage{url}
\usepackage[colorlinks = true,
            linkcolor = black,
            urlcolor  = black,
            citecolor = black,
            anchorcolor = blue]{hyperref}

\newcommand{\djh}[1]{\textcolor{black}{#1}}

\begin{document}

\title{Non-reciprocal circular dichroism of axial phonons coupled to ferro-rotational order}

\author{H. Y. Huang}
\affiliation{National Synchrotron Radiation Research Center, Hsinchu 300092, Taiwan}

\author{G. Channagowdra}
\affiliation{National Synchrotron Radiation Research Center, Hsinchu 300092, Taiwan}

\author{D. Banerjee}
\affiliation{Department of Physics and Astronomy, The University of Tennessee, Knoxville, Tennessee 37996, USA}
\affiliation{Institute for Advanced Materials and Manufacturing, University of Tennessee, Knoxville, Tennessee 37996, USA}

\author{E. V. Komleva}
\affiliation{Institute of Metal Physics, 620041 Ekaterinburg GSP-170, Russia}
\affiliation{Department of Theoretical Physics and Applied Mathematics, Ural Federal University, 620002 Ekaterinburg, Russia}

\author{J. Okamoto}
\affiliation{National Synchrotron Radiation Research Center, Hsinchu 300092, Taiwan}

\author{C. T. Chen}
\affiliation{National Synchrotron Radiation Research Center, Hsinchu 300092, Taiwan}

\author{M. Guennou}
\affiliation{Department of Physics and Materials Science, University of Luxembourg, 4362 Esch-sur-Alzette, Luxembourg}

\author{S. Johnston}
\affiliation{Department of Physics and Astronomy, The University of Tennessee, Knoxville, Tennessee 37996, USA}
\affiliation{Institute for Advanced Materials and Manufacturing, University of Tennessee, Knoxville, Tennessee 37996, USA}

\author{S. V. Streltsov}
\affiliation{Institute of Metal Physics, 620041 Ekaterinburg GSP-170, Russia}
\affiliation{Department of Theoretical Physics and Applied Mathematics, Ural Federal University, 620002 Ekaterinburg, Russia}

\author{C. Y. Mou}
\affiliation{Department of Physics, National Tsing Hua University, Hsinchu 300044, Taiwan}
\affiliation{Center for Quantum Science and Technology and Department of Physics, National Tsing Hua University, Hsinchu 300044, Taiwan}

\author{A. Fujimori}
\affiliation{National Synchrotron Radiation Research Center, Hsinchu 300092, Taiwan}
\affiliation{Department of Physics, National Tsing Hua University, Hsinchu 300044, Taiwan}
\affiliation{Center for Quantum Science and Technology and Department of Physics, National Tsing Hua University, Hsinchu 300044, Taiwan}
\affiliation{Department of Physics, University of Tokyo, Bunkyo-Ku, Tokyo 113-0033, Japan}

\author{S.-W. Cheong}
\email{sangc@physics.rutgers.edu}
\affiliation{Keck Center for Quantum Magnetism and Department of Physics and Astronomy, Rutgers University, Piscataway, NJ 08854, USA}

\author{D. J. Huang}
\email{djhuang@nsrrc.org.tw}
\affiliation{National Synchrotron Radiation Research Center, Hsinchu 300092, Taiwan}
\affiliation{Department of Physics, National Tsing Hua University, Hsinchu 300044, Taiwan}
\affiliation{Department of Electrophysics, National Yang Ming Chiao Tung University, Hsinchu 300093, Taiwan}

\date{\today}

\begin{abstract}
\djh{Circular dichroism (CD) in X-ray absorption, defined as the difference in absorption between opposite circular polarizations, is fundamentally enabled by the breaking of time-reversal symmetry or inversion symmetry. It is therefore sensitive to magnetism, chirality, and their interplay. In contrast, the sample symmetry alone is insufficient to determine whether CD in resonant inelastic X-ray scattering (RIXS) is allowed. Rather, RIXS-CD is governed by both the sample symmetry and the scattering geometry. Here, using RIXS,} we identify circularly polarized phonons coupled to ferro-rotational order in MnTiO$_3$, which we refer to as ferro-axial phonons. Their excitations provide a direct demonstration of non-reciprocal \djh{RIXS-CD, in which the dichroic response changes upon reversing the propagation direction of the incident X-rays, although the system globally preserves both inversion and time-reversal symmetries.} We propose that a condensate of these phonons, manifested as standing waves, underlies the ferro-rotational order in MnTiO$_3$. The observed non-reciprocal CD reflects the interplay among photon helicity, phonon polarization, and ferro-rotational order. 
\end{abstract}

\maketitle
\section{Introduction}
Symmetry plays a fundamental role in condensed matter physics, and its breaking often signals a phase transition and the emergence of an ordered state. X-ray circular dichroism (CD), defined as the difference in absorption or scattering intensity between left- and right-handed circularly polarized (LCP and RCP) X-rays, provides a powerful probe of symmetry breaking in solids~\cite{Alagna1998,Goulon1998,Goulon2003, Oreshko2018,okamoto2024giant}. \djh{
CD in X-ray absorption requires broken time-reversal symmetry for X-ray magnetic circular dichroism and broken inversion symmetry for X-ray natural circular dichroism. It is therefore sensitive to magnetism, structural chirality, as well as their interplay. In contrast, the sample symmetry alone is insufficient to determine whether circular dichroism in resonant inelastic X-ray scattering (RIXS-CD) is allowed. Rather, RIXS-CD is governed by the Hamiltonian's unitary symmetries together with the scattering geometry, rather than directly by time-reversal symmetry \cite{Furo2025PRB}. If a unitary symmetry, such as a mirror operation, relates the two incident helicities while leaving the RIXS measurement invariant, the CD must vanish. Thus, the scattering geometry is an essential component of the symmetry analysis. Furthermore, reciprocal RIXS-CD such as that of chiral phonons~\cite{Ueda2023} satisfies the reciprocity relation, yielding identical circular dichroism for opposite incident X-ray wavevectors. In contrast, non-reciprocal RIXS-CD, if observed, would violate this relation, producing different dichroic responses for opposite incident wavevectors $\pm\bf{k}$. The reciprocity relation therefore provides an additional symmetry criterion for characterizing the underlying excitations.}


\djh{Guided by this symmetry framework}, one can explore materials with unconventional order parameters. In ferromagnets, spin alignment breaks time-reversal symmetry, while ferroelectrics exhibit a spontaneous polarization $\bf{P}$ that breaks inversion symmetry. Ferro-rotational materials, by contrast, possess a finite and uniform axial electric toroidal moment density, ${\bm{\mathcal A}}\equiv\sum_{i}{\bf r}_{i}\cross{\bf p}_{i}$, where ${\bf r}_{i}$ denotes the position vector of electric dipole ${\bf p}_{i}$~\cite{Gopalan2011rotation, Johnson2012_PRL, Cheong2018broken,Jin2020_NP,Hayashida2020}. $\bm{\mathcal{A}}$ is an axial vector that governs the rotational electric-dipole arrangement. Also called ferroaxial order, ferro-rotational order describes the presence of rotational distortions and remains invariant under both time-reversal and spatial-inversion operations. Although relatively rare, ferro-rotational order occurs in complex oxides with uniform rotations of oxygen cages~\cite{Hayashida2020,yokota2022,kidoh1984,Hayashida2021,Sekine2024,Bhowal2024,kusuno-prl,zhang2025}, \djh{including ilmenite MnTiO$_3$ and NiTiO$_3$, which are centrosymmetric and possess a with a well-defined axial order parameter~\cite{kidoh1984,Hayashida2021,Sekine2024,Bhowal2024,kusuno-prl}. This unique symmetry provides an ideal platform for exploring the interplay between axial order and circular lattice dynamics, for which circularly polarized phonons provide a natural realization.}

In general, chiral systems possess a well-defined handedness arising from the absence of mirror and inversion symmetries. Chiral phonons, originating from the rotational motion of atoms in the lattice, carry pseudo-angular momentum and are associated with a variety of novel collective phenomena~\cite{Zhang_chiral_phonon,kishine2020,Zhu2018,YinAdvMater2021,Ishito2023,Ueda2023,okamoto2025,juraschek2025}. Because of symmetry constraints, intrinsic chiral phonons with symmetry-protected handedness require the absence of inversion symmetry. Interestingly, recent studies have revealed planar chiral phonons in two-dimensional systems~\cite{Zhu2018}. Planar chirality arises when reflection symmetry is broken with respect to two orthogonal mirror planes rather than all mirror planes, while inversion symmetry is preserved. Consequently, although planar chiral phonons exhibit circular lattice motion, 
they are not truly chiral phonons in the symmetry sense. Whether an axial order parameter can support a distinct class of circularly polarized phonons while preserving inversion symmetry remains unexplored.


In this Article, we report the observation of circularly polarized phonons in ferro-rotational MnTiO$_3$ in both paramagnetic and antiferromagnetic phases using RIXS with circularly polarized light. RIXS is a powerful spectroscopic technique for studying collective excitations in solids~\cite{Mitrano2024exploring}, including phonons and their coupling to electrons~\cite{Ament11a, Lee2013role, Devereaux2016, Rossi2019experimental, Geondzhian2020generalization, huang2021quantum, Ueda2023, thomas2024theory}. Together with exact diagonalization (ED) and density functional theory (DFT) calculations, we identify orbital-selective phonon excitations involving the hybridization of O $2p$ states with Ti $3d$ states. Two-component $E_g$ phonon modes exhibit circular dichroism, manifesting as a clear contrast in RIXS spectral weight between LCP and RCP X-rays. This circular dichroism reverses sign upon reversing the incident wavevector $\bf{k}$, revealing a non-reciprocal response that signifies the presence of circularly polarized phonons. We further show that these phonon excitations, originating from rotational lattice distortions, are coupled to 
ferro-rotational order, even though the local chirality is globally canceled in MnTiO$_3$ due to its inversion symmetry. As circularly polarized phonons are axial in nature, we term these excitations ferro-axial phonons \cite{Martinez2025,Wang2025alteraxial}.

\begin{figure*}[ht!]
\centering
\includegraphics[width=0.9\linewidth]{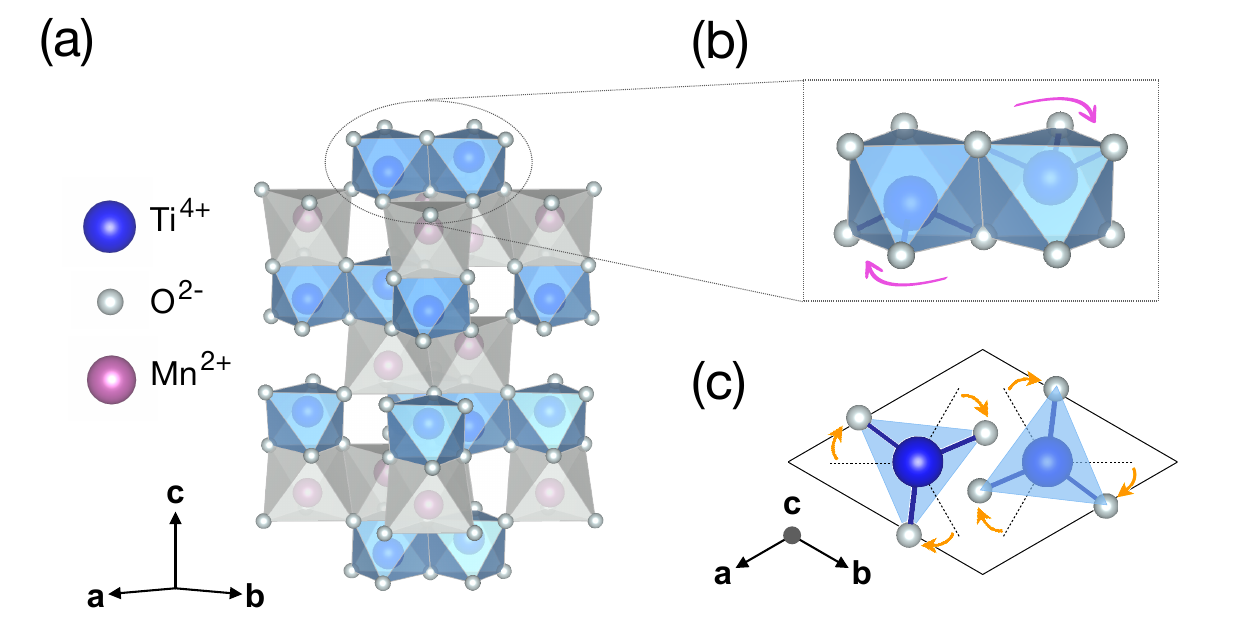}
\caption{{\bf Crystal Structure of ilmenite MnTiO$_3$.} (a) The crystal structure of MnTiO$_3$ in the trigonal space group R$\bar{3}$. (b) An enlarged view of two trigonally distorted TiO$_6$ octahedra, with purple arrows indicating the direction of the distortional twist. Blue bonds highlight the connections between Ti ions and their nearest oxygen neighbors, forming triangular pyramids. (c) A view along the $c$-axis of the two TiO$_3$ pyramids shown in (b). Orange arrows denote the rotational displacements of oxygen ions relative to the (110)-type planes (dotted lines). }\label{fig:crystal}
\end{figure*}

\section{RIXS Results}

We detected MnTiO$_3$ phonon excitations using O $K$-edge RIXS with circularly polarized incident X-rays, without polarization analysis of scattered X-rays. MnTiO$_3$ crystallizes in the ilmenite structure with space group $R\bar{3}$ (No.~148), whose centrosymmetric point group $\bar{3}$ ($C_{3i}\equiv C_3\otimes C_i$) prohibits macroscopic chirality but allows ferro-rotational order. The lattice consists of alternating layers of edge-sharing MnO$_6$ and TiO$_6$ octahedra stacked along the hexagonal $c$-axis, as shown in Fig.~1(a). 
Within the ilmenite structure, the TiO$_6$ and MnO$_6$ octahedra exhibit rotational distortions of opposite sense about the $c$-axis. Both octahedra exhibit trigonal crystal-field distortions, and in combination these effects lead to oxygen-ion rotational displacements, as illustrated in Fig.~\ref{fig:crystal}(b). The resulting local chiralities cancel globally, conforming to the inversion symmetry of the ilmenite structure. Nevertheless, the collective octahedral rotations induce electric dipole moments arranged in a circular pattern. These rotational lattice distortions give rise to a ferro-rotational order characterized by an electric-toroidal moment along the $c$-axis, as shown in Fig.~\ref{fig:crystal}(c). In addition, MnTiO$_3$ is a uniaxial antiferromagnet with a N\'eel temperature $T_\mathrm{N}\sim 64$~K, where Mn$^{2+}$ moments order in a collinear $G$-type antiferromagnetic pattern aligned along the $c$-axis. It exhibits linear magnetoelectric coupling~\cite{Mufti2011}, similar to that observed in Cr$_2$O$_3$.

At the O $K$-edge, the incoming photon excites an electron from the oxygen $1s$ core orbital into a higher-energy unoccupied 
state involving the O $2p$ orbital hybridized with the Ti or Mn $3d$ orbitals. This excitation generates a transient intermediate state composed of a core hole and an electron in an excited state. The intermediate state interacts with the surrounding lattice through the electron-phonon ($e$-ph) coupling, generating phonon excitations. In the subsequent step, another electron in the excited O $2p$ states relaxes back to fill the original $1s$ core hole, emitting a photon in the process. By tuning the energy of incident X-ray at the O $K$-edge to the $2p$ hybridized with a specific Ti or Mn $3d$ state,  phonon excitation associated with the corresponding orbital can be selectively probed, as illustrated by Fig.~S1 in the Supplemental Material. Phonons in MnTiO$_3$ reflect locally chiral distortions of the octahedra: in particular, $E_g$ phonon modes involving rotational motion of the TiO$_6$ cages~\cite{Lujan2024, kusuno-prl} display circular dichroism in RIXS. These polarized phonon excitations provide a spectroscopic fingerprint of the ferro-rotational order and the magnetoelectric ground state in MnTiO$_3$.

\begin{figure*}[t]
\centering 
\includegraphics[width=1.8\columnwidth]{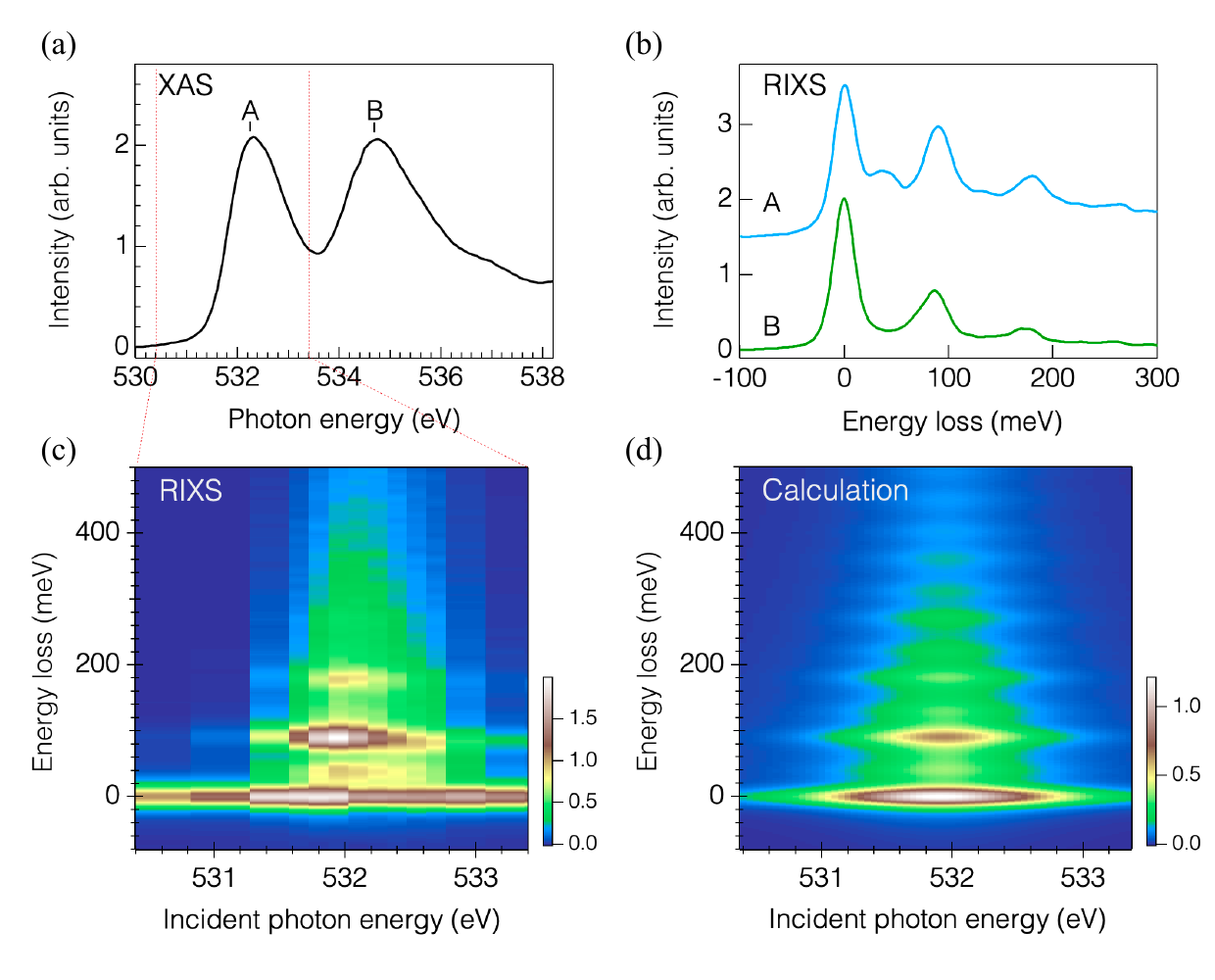}
\caption{{\bf O $K$-edge XAS and RIXS of MnTiO$_3$.} (a) XAS spectrum recorded in the fluorescence yield mode. 
(b) RIXS spectra measured with left-handed circularly polarized (LCP) X-rays at 
two incident X-ray energies, denoted by A and B, corresponding to the transition from O $1s$ to $2p$ states hybridized with Ti $t_{2g}$ and $e_g$ states, respectively.
(c) RIXS intensity map plotted in the plane of energy loss vs. incident X-ray energy. 
The scattering angle was 150\textdegree, and the incidence angle was 75\textdegree. The corresponding momentum transfer is $|\mathbf{q}| \approx 0.52~\AA^{-1}$ along the $(0,0,1)$ direction. All data were measured at 150~K.  (d) Theoretical RIXS intensity map simulated by using a two-mode model.}
\label{fig:3}
\end{figure*}

Figure~\ref{fig:3}(a) shows O $K$-edge XAS results for MnTiO$_3$. The spectrum  exhibits two broad pre-edge features separated by 2.3~eV, which we denote as A and B. Our DFT+$U$ calculations shown in Fig. S2 in the Supplemental Material  reveal that the first peak (A) corresponds to transitions from the $1s$ state to $2p$ states hybridized with Ti $t_{2g}$ states, while the second peak (B)  corresponds to transitions to states hybridized with Ti $e_g$ or Mn $t_{2g}$ states, consistent with previous results \cite{ali2024}. Figure~\ref{fig:3}(b) plots the RIXS spectra at O $K$-edge with incident photons tuned to these two XAS peaks. When the incident photon energy is set to peak A, two phonon excitations appear at $\Omega \approx 34$ and 90 meV, along with overtones of the 90-meV phonon. Notably, the 90 meV mode is also present for incident energies tuned to the B peak, whereas the 34-meV phonon excitation is absent. This behavior indicates that excitation of the lower-energy mode is orbital-selective in nature~\cite{Lee2013role} and is excited only in conjunction with excitations into the Ti $t_{2g}$ states. To further examine these phonon excitations, Fig.~\ref{fig:3}(c) 
plots a RIXS intensity map plotted as a function of incident X-ray energy within a selected energy window covering peak A in the XAS. 
For comparison, Fig.~\ref{fig:3}(d) shows a corresponding modeled spectra obtained from fitting RIXS data at the A resonance with a multi-mode atomic model~\cite{Geondzhian2020generalization} (see the Supplemental Material).

\begin{figure*}[t]
\centering 
\includegraphics[width=1.8\columnwidth]{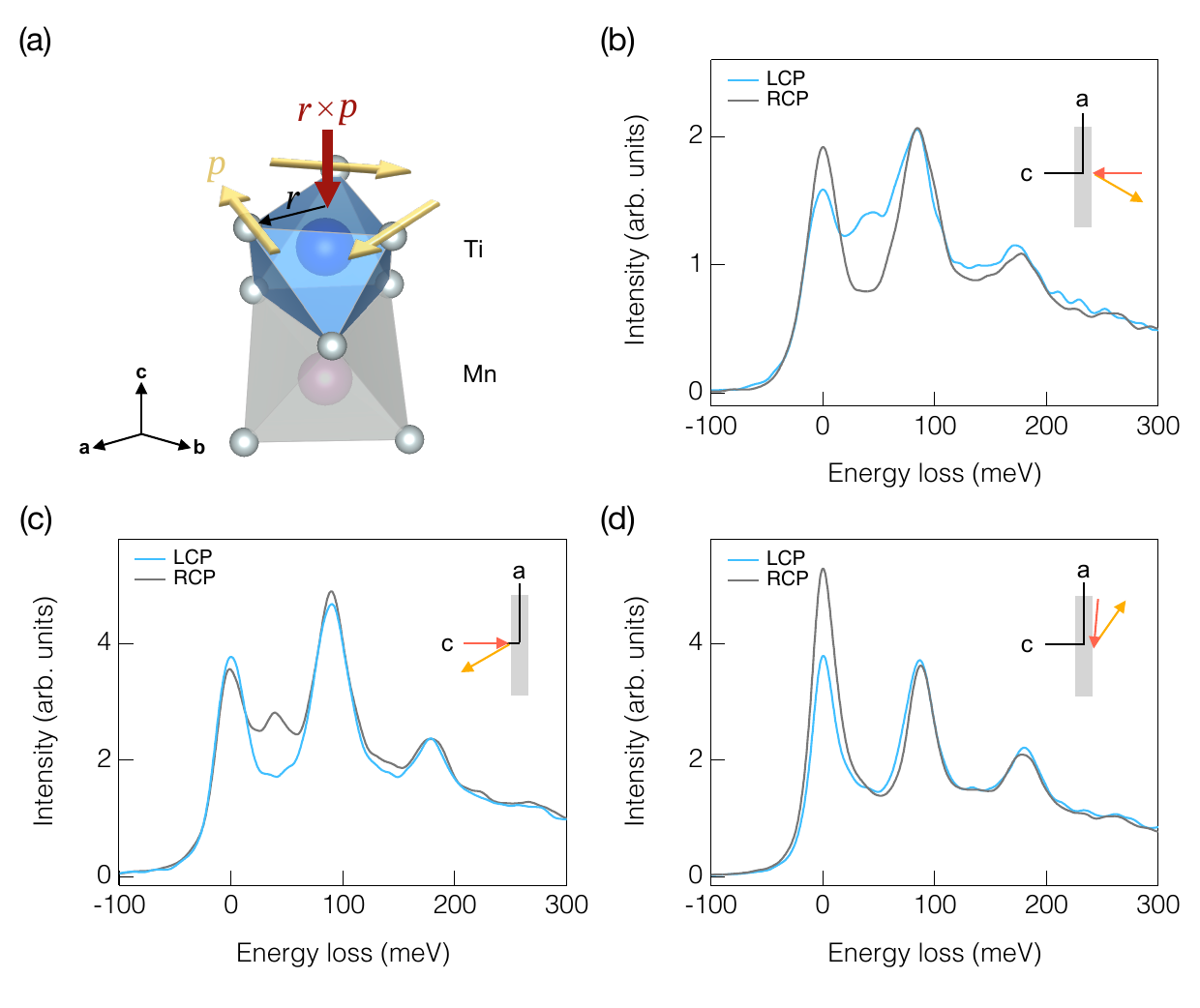}
\caption{{\bf Ferro-axial phonons in MnTiO$_3$ observed by RIXS.} (a) Cartoon illustration of the sample orientation in relation to its ferro-rotational moment. The sample surface is parallel to the $ab$ plane, with the axial vector of the ferro-rotational moment along the crystal $c$-axis. (b)-(d) RIXS spectra measured with circularly polarized X-rays at the incident photon energy A indicated in Fig. 2(a) under various conditions. RCP and LCP denote right- and left-handed circularly polarized incident X-rays, respectively. All data were measured at 150~K. The directions of the incident and scattered X-rays ($\omega_{\rm in}$ and $\omega_{\rm out}$) are indicated by red and orange arrows, respectively. The RIXS scattering angle, defined as the angle between the incident and scattered X-rays, was set to 150$^\circ$, corresponding to a momentum transfer of magnitude $|\mathbf{q}| \approx 0.52~\AA^{-1}$. The spectra in (b) and (c) were taken with X-rays propagating in opposite directions along the $c$-axis. The spectra plotted in (d) were recorded with an X-ray incident angle of 10$^\circ$. Planar chiral phonons of energy 34 meV appear only in RIXS measurements with X-rays normally incident to the sample $ab$ plane.}\label{fig_3}
\end{figure*}

To clarify the nature of the 34-meV phonon, Fig.~\ref{fig_3} analyzes its circular dichroism in the paramagnetic phase. Measurements with LCP and RCP X-rays reveal a clear contrast in the excitations of polarized phonons by opposite photon helicities. Switching the X-ray polarization from LCP to RCP transfers an angular momentum of ${\pm}2\hbar$ to phononic excitations, as discussed below. The resulting intensity difference between the RCP and LCP spectra exceeds 80\% and indicates a polarized phonon mode. Residual asymmetries from X-ray birefringence are expected to be small under normal-incidence conditions. The large circular dichroism suggests that the sample is close to a single-domain state. In contrast, the higher-energy phonon and its overtone show no dichroism, implying that the RIXS-CD of the 34-meV phonon is intrinsic. Moreover, the observed phonon dichroism reverses sign upon flipping the incident X-ray direction, as shown in Figs.~\ref{fig_3}(b) and \ref{fig_3}(c), demonstrating its non-reciprocal character. Furthermore, a grazing-incidence X-ray does not activate the 34 meV polarized phonon, indicating that its excitation occurs at a vanishing or small in-plane momentum transfer $\bf{q_\|}$,  as shown in Fig.~\ref{fig_3}(d). This phonon mode \djh{also shows negligible dispersion along the $\Gamma{A}$ direction as shown in Fig.~S3 in the Supplemental Material. It}, therefore, corresponds to finite-$q_z$ excitations near the two-dimensional Brillouin-zone center. Notably, the circular dichroism in the 34-meV phonon described above remains unchanged below $T_\mathrm{N}$ in the antiferromagnetic phase (see Fig.~S4 in the Supplemental Material). Together, these results suggest that the observed axial phonons are located at the two-dimensional zone center and are not of magnetic origin. From a symmetry perspective, these planar phonons are achiral and therefore do not constitute truly chiral phonons.

\section {Discussion}

\subsection{Orbital-selective electron-phonon coupling}
The incident-energy–dependent RIXS measurements indicate that the 34~meV low-energy phonon mode primarily involves the Ti $t_{2g}$ orbitals rather than the $e_g$ orbitals. The RIXS intermediate state is therefore dominated by $\pi$-type ($pd\pi$) overlap between O $2p$ and Ti $3d$ orbitals, mediated by the components of the O $2p$ orbitals that are perpendicular to the Ti–O bond direction. Consistently, our DFT calculations show that the 34 meV mode is
characterized by O displacements largely transverse to the Ti–O bonds, whereas the 90-meV mode involves displacements along the Ti–O bonds. Although weaker in RIXS intensity, the 34 meV mode originates from bond-bending or octahedral-rotation motions and exhibits a strong effective $e$-ph coupling from the $E_g$ phonon.

\begin{figure}[t]
\centering 
\includegraphics[width=0.9\columnwidth]{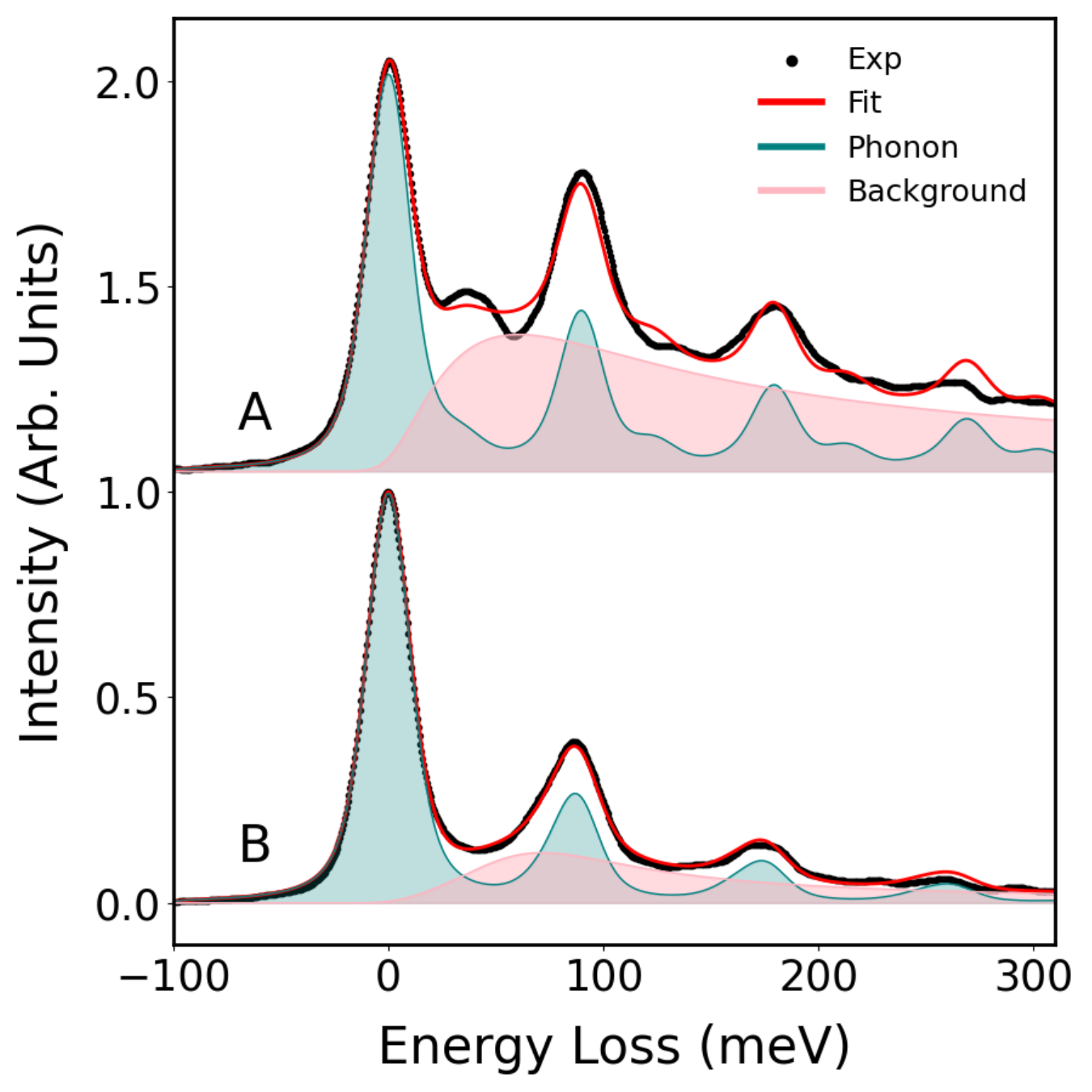}
\caption{{\bf Fit results for the RIXS spectra measured at each resonance}. 
The black dots indicate the experimental data, and the red curves are the results for the fitted spectra. The green and pink shaded curves show the contributions to the fit from the additional two phonon modes and the background continuum, respectively. 
Table~\ref{tbl:fits} provides the 
all fit parameters and estimated error bars. All spectra were normalized to the intensity of the elastic line.}
\label{fig:phonon_fits}
\end{figure}

To assess the relative coupling to the phonons probed at each resonance, we fit the RIXS spectra shown in Fig.~\ref{fig:3}(c) using a generalized atomic model with coupling to two phonon modes~\cite{Geondzhian2020generalization} supplemented with an additional background contribution (see the Supplemental Material), as shown in Fig.~\ref{fig:phonon_fits}. The results confirm what is already evident in the raw data. The A resonance has features consistent with strong coupling to a local $\Omega \approx 89.66$ and $\Omega \approx 34.01$ meV modes. (The precise values obtained from the fits and estimated error bars are provided in Table~\ref{tbl:fits}.) Conversely, the B resonance has features arising from coupling to an $\Omega \approx 87.92$ mode with a weaker secondary coupling to an $\Omega \approx 73.06$ meV mode. The high-energy mode is consistent with a non-chiral transition metal-oxygen bond stretching phonon and appears in both resonances. It, therefore, likely originates from the same, possibly weakly dispersing, phonon branch. The fitting results suggest that the strength of coupling to this mode depends on whether the core electron is excited into the Ti or Mn $3d$ states, analogous to prior observations of orbital-dependent coupling strengths in spin chain cuprates~\cite{Lee2013role}. Conversely, the coupling to the $\Omega \approx 34.01$ mode 
is clearly orbital selective; it only becomes active for incident photon energies tuned to the $A$ resonance when the core electron is excited into the Ti $t_{2g}$ orbitals. 

\begin{table}[h]
\centering
\caption{Best fit model parameters for the modeling shown in Fig.~\ref{fig:phonon_fits}.}\label{tbl:fits}
\begin{tabular}{lll}
\hline
\hline
\multicolumn{1}{c}{\textbf{Parameter}} & \multicolumn{1}{c}{\textbf{Resonance A} (meV)} & \multicolumn{1}{c}{\textbf{Resonance B} (meV)} \\ \hline
$\Omega_1$  & $34.017 \pm1.8$ & $87.92 \pm 0.15$\\    
$\Omega_2$  & $89.66 \pm 0.3$ & $73.06\pm 0.74$\\    
$M_1$ & $104.47 \pm 3.75$ & $114.44 \pm 0.72$\\
$M_2$ & $284.41 \pm 2.04$ & $48.69 \pm1.34$\\
$A_{e-\text{ph}}$ & $1403.9 \pm 14.91$ & $2470.7 \pm 25.63$\\
$A_{\text{back}}$ & $17267.1 \pm 183.1$ & $9726.4 \pm 100.8$ \\
$\Omega_\text{back.}$ & $-0.472 \pm 0.76$ & $53.9 \pm 0.68$ \\
$\gamma_\text{back.}$ & $59.18 \pm0.86$ & $46.6 \pm 0.38$ \\
\hline
\hline
\end{tabular}
\end{table}

\subsection{Ferro-axial phonons}
The low-energy phonon mode observed in the O $K$-edge RIXS spectra of MnTiO$_3$, excited by circularly polarized X-rays, is characterized by strong non-reciprocal circular dichroism, and appears only for an in-plane momentum transfer $\bf{q}_{\|}$ close to zero. 

For a crystal with $C_{3i}$ symmetry about the $c$-axis (taken as the $z$-axis), the irreducible representations of phonons at the zone center $\Gamma$ are $A_g$, $E_{g}$, $A_u$, and $E_{u}$,  where $A$ and $E$ denote one- and two-dimensional irreducible representations, respectively, and 
the subscript $g$ ($u$) indicates even (odd) parity. (See Table S1 in the Supplemental Material  for the character table of the $C_{3i}$ point group.) $E_g$ phonons at the $\Gamma$ point can contain circularly polarized modes $E^{\pm}_{g}$ that carry finite angular momentum along the $c$-axis without violating inversion symmetry~\cite{Coh2023,Tang2024,Mustafa2025}. 
The observed phonons correspond to finite-$q_z$ excitations near $\Gamma$. Although the $E_g^\pm$ basis is exact only at the zone center, it remains a good approximation \djh{for phonon modes with negligible dispersion along} $q_z$, where the two transverse branches derived from the $E_g$ doublet remain nearly degenerate, \djh{thereby preserving} a well-defined phonon rotation direction~\cite{Zhu2018}.

Our DFT+$U$ calculations 
show that the $E_g^{\pm}$ phonons possess opposite circular polarizations. The calculations identify five $E_g^{\pm}$ phonon doublets, two of which exhibit strong circular polarization and correspond to $E_g^{\pm}$ modes with predominantly Mn and Ti character (see Fig.~S5 in the Supplemental Material). Because the incident photon energy is tuned to O $2p$ states hybridized with the Ti $t_{2g}$ orbitals, our O $K$-edge RIXS primarily probes the $E_g^{\pm}$ phonons associated with the Ti $t_{2g}$ states and hence of Ti character. 
From a symmetry perspective, the $E^{\pm}_{g}$ phonons are not chiral. Our RIXS measurements with circularly polarized X-rays under different scattering geometries indicate that the 34-meV phonon peak arises from circularly polarized phonons carrying a finite angular momentum along the $c$-axis. 
For incident and scattered X-rays nearly along the $c$-axis, an incident X-ray with RCP and LCP excites $E^{+}_{g}$ and $E^{-}_{g}$ phonons, respectively. In a $C_{3i}$ crystal, symmetry requires that the change in angular momentum be a multiple of $3\hbar$; that is, the pseudo-angular momentum is defined modulo $3\hbar$ and can take the three eigenvalues 0 and $\pm\hbar$, corresponding to the $A$ and $E$ representations, respectively. Consequently, an angular momentum of $\pm 2\hbar$ transferred by the X-ray scattering process~\cite{Ueda2023,okamoto2025} corresponds to a pseudo-angular momentum of $\mp\hbar$. The angular momenta of circularly polarized $E_g^{\pm}$ phonon are therefore $\mp \hbar$.  Consequently, the circular-dichroism signal appears if the crystal hosts a single ferro-rotational domain, and it reverses sign when the incident beam direction is flipped, reflecting the relative alignment of phonon angular momentum and ferro-rotational (ferroaxial) moment.

\begin{table*}[t] 
    \centering
    \caption{Symmetry assignment of the RIXS phonon peaks in MnTiO$_3$ under $C_{3i}$ symmetry. Circular polarizations RCP and LCP are defined by the helicity of the incident or scattered X-rays, whereas the $E^{\pm}_u$ symmetries are defined with respect to the crystallographic axes. The energy assignment is based on the RIXS data, in which the 34-meV phonons exhibit strong circular dichroism, whereas the 90-meV phonons show no circular dichroism. Measured RIXS spectra with LCP and RCP X-rays indicate $I_2{\gg}I_1$, consistent with the expected intensities from a sample with a single ferro-rotational domain.}
    \begin{tabular}{c c c c c c c c}
\\[0.5mm]
\hline
\hline
Incident X-ray&\multicolumn {2}{c}{Incident X-ray}& \multicolumn {2}{c}{Scattered X-ray} &  \multicolumn {2}{c}{Phonon} & RIXS  \\
Direction& Polarization&Symmetry&Polarization &Symmetry&Symmetry &Energy
&Intensity\\
   \hline
   \hline
   \\[0.1mm]
 Front face  & RCP  & $E^{+}_{u}$ & RCP & $E^{-}_{u}$ & $E^{-}_{g}$ & 34 meV
 & $I_2$ \\
   &  && LCP  & $E^{+}_{u}$ & $A_g$ & 90 meV
   &  \\
   \hline
   \\[0.1mm]
   & LCP  & $E^{-}_{u}$ & RCP &$E^{-}_{u}$& $A_g$ & 90 meV
   &  \\
   &  && LCP &$E^{+}_{u}$& $E^{+}_{g}$ & 34 meV
   & $I_1$ \\
 \hline
 \hline
 \\[0.1mm]
Back face  & RCP & $E^{-}_{u}$&RCP& $E^{+}_{u}$& $E^{+}_{g}$ & 34 meV 
& $I_1$ \\
&  && LCP& $E^{-}_{u}$& $A_g$ & 90 meV 
& \\
   \hline
   \\[0.1mm]
   & LCP &$E^{+}_{u}$& RCP & $E^{+}_{u}$& $A_g$  & 90 meV 
   & \\
   && & LCP& $E^{-}_{u}$& $E^{-}_{g}$ & 34 meV
   & $I_2$\\
\hline
\hline
\end{tabular}
\label{rixs-cd_c3}
\end{table*}

Table~\ref{rixs-cd_c3} summarizes the above observations and symmetry analysis for the two incident beam directions, namely, incident on the front and back faces of the sample. Under the $C_{3i}$ symmetry, the energies of the $E^{+}_{g}$ and $E^{-}_{g}$ phonons are degenerate, while their intensities are generally different~\cite{kusuno-prl}, as denoted by $I_1$ and $I_2$ in Table~\ref{rixs-cd_c3}, respectively.   
Since the circular polarization of scattered X-rays was not analyzed in the experiment, RIXS spectra for each incident X-ray direction and each circular polarization are given by summing over the circular polarization of scattered X-rays, that is, by summing two lines in each segment of the table.     
Along this line, the RIXS circular dichroism data indicate one of the two 34-meV modes with significantly high intensity, i.e., $I_2 \gg I_1$, showing our MnTiO$_3$ crystal exhibits a single ferro-rotational domain.
When $\bf{q}$ is oriented nearly perpendicular to the $c$-axis, both the 34-meV phonon intensity and its dichroism vanish. This behavior indicates that low-energy phonons carrying finite angular momentum can be excited only at $\bf{q}_{\parallel}=0$, consistent with the symmetry requirement for $E_g$ phonons to acquire angular momentum. These symmetry considerations establish the 34 meV excitation as a circularly polarized phonon mode whose angular momentum is intimately tied to the ferro-rotational order along the crystallographic $c$-axis. The observed ferro-axial phonons may serve as a dynamical signature of ferro-rotational order, analogous to how magnons reflect ferromagnetic order.

\subsection{Non-reciprocal circular dichroism}
In a centrosymmetric and time-reversal-symmetric crystal, the \djh{X-ray absorption} cross sections are generally expected to be insensitive to the handedness of X-ray circular polarization. Such systems lack the pseudoscalar quantities required to generate circular dichroism in optical or X-ray spectroscopies. The observation of non-reciprocal RIXS-CD in the phonon excitations of MnTiO$_3$ in its paramagnetic phase is therefore highly nontrivial. 
\djh{This observation alone does not imply any symmetry breaking of the phonon. Instead, its non-reciprocity indicates that the RIXS scattering amplitude contains an axial-pseudoscalar contribution permitted by the ferro-rotational symmetry. Our results demonstrate that a finite ferro-rotational order parameter is selectively coupled to the X-ray helicity, as elaborated below.} 

The cross section of RIXS is determined by the inelastic scattering amplitude $F$, which is governed by the interaction that scatters the incident X-ray photon with wavevector $\bf{k}$ into an outgoing photon with wavevector $\bf{k}'$, accompanied by the creation of a ferro-axial phonon. To identify the relevant interaction, we note that an electric field $\bf{E}$ moving with velocity $\bf{v}$ generates a magnetic field $\bf{B}\!\propto\!\bf{E}\!\times\!\bm{v}$. 
This magnetic field can interact with spin $\bf{S}$ through a Zeeman-type interaction $-\bf{B}\cdot\bf{S}$, which gives rise to the standard spin-orbit interaction. By the same reasoning, the electric field of circular X-ray photon can be represented by the polarization  $\bm{\epsilon}_{\pm}=\bm{\epsilon}_x{\pm}{i}\bm{\epsilon}_y$, 
the atom velocity can be represented by the phonon polarization  $\bm{\epsilon}_{\pm}^{q}=\bm{\epsilon}_x^{q}{\pm}i\bm{\epsilon}^{q}_y$, 
and the spin can be represented by the axial order parameter $i\omega\bm{\mathcal{A}}$ as $i{\omega}\bf{p}$ being proportional to atom momentum. Here $\omega$ is frequency of phonon. For the coupling between the photon field and the phonon mode in the presence of a ferroaxial order parameter $\bm{\mathcal A}$, symmetry considerations require that the lowest-order term in the RIXS scattering amplitude takes the form $\bm{\epsilon} \times \bm{\epsilon}^{q} \cdot \bm{\mathcal A}$.  From a microscopic perspective, as detailed in the Supplemental Material, a more complete expression of the scattering amplitude is
$F = F_0 + \lambda_0(\omega, \omega')\, \bm{\epsilon} \times \bm{\epsilon}^{q} \cdot \boldsymbol{\mathcal A},$
where $F_0$ is the helicity-independent contribution and $\lambda_0(\omega, \omega')$ is the coupling strength, whose magnitude can be evaluated within  a microscopic model. 
Note that $\bf{\epsilon}$ and $\bf{\epsilon}^{q}$ represent two angular momenta and change sign under time-reversal transformation. As a result, the RIXS intensity
can be expressed as 
\begin{equation}
I \propto |F_0|^2+2\mathrm{Re}\left[{\lambda}(\omega, \omega')\bm{\epsilon}\times\bm{\epsilon}^{q}\cdot\boldsymbol{\mathcal{A}}\right]+ \mathcal{O}(\lambda^2),
\label{I_eXeA}
\end{equation}
where $\lambda \equiv F_0^{*}\lambda_0(\omega, \omega')$. The non-reciprocal signal observed in RIXS therefore originates from the interference of the scattering amplitudes $F_0$ and $\bm{\epsilon}\times\bm{\epsilon}^{q}\cdot\boldsymbol{\mathcal{A}}$. This reflects the interplay of photon polarization, 
phonon polarization, and ferroaxiality~\cite{He_2024}. The coupling term $\bm{\epsilon}\times \bm{\epsilon}^{q}\cdot \mathbf{\mathcal A}$ is odd under reversal of the incident photon direction\footnote{For incident X-rays propagating along $+\hat{\mathbf z}$, the circular polarization vectors are $\bm{\epsilon}_{\pm}=\tfrac{1}{\sqrt{2}}(\hat{\mathbf x}\pm i\hat{\mathbf y})$. In this basis, the $z$-component of the cross product vanishes for the same rotation direction, $(\bm{\epsilon}_{\pm}\times \bm{\epsilon}^{q}_{\pm})_z = 0$,
so the coupling $(\bm{\epsilon}\times \bm{\epsilon}^{q})_z$ is nonzero only for opposite photon and phonon rotation directions. Upon reversing the incident and scattered X-ray directions along the $z$-axis, the circular polarization of photon changes as $\bm{\epsilon}_{\pm}\rightarrow \bm{\epsilon}_{\mp}$, and the corresponding axial phonon components are interchanged, $\bm{\epsilon}^{q}_{\pm}\rightarrow \bm{\epsilon}^{q}_{\mp}$. Hence, the cross product changes sign, $(\bm{\epsilon}_{+}\times \bm{\epsilon}^{q}_{-})_z = -(\bm{\epsilon}_{-}\times \bm{\epsilon}^{q}_{+})_z,$ leading to a sign reversal of $(\bm{\epsilon}\times \bm{\epsilon}^{q})_z$, while the ferro-rotational moment density $\bm{\mathcal A}$ remains unchanged}, whereby the circular dichroism reverses sign. This model thus explains the observed nonreciprocity between Figs. 3(b) and (c). 

More generally, the expansion in Eq.~(\ref{I_eXeA}) follows from the coherent and dynamical nature of the RIXS process: the measured intensity is governed by a frequency-dependent scattering amplitude, in which an axial order parameter enters and contributes linearly to the intensity through interference with the dominant helicity-independent channel. The essential requirement for observing non-reciprocal circular dichroism is therefore the presence of a finite axial order parameter in the system. 
Motivated by this observation, we propose a scenario in which the ferro-rotational order arises from the condensation of ferro-axial phonons. Pairs of counter-propagating circularly polarized phonons in MnTiO$_3$ give rise to a rotational motion of atoms with a well-defined pseudo-angular momentum, for example $L_z = \mp\hbar$, whose degeneracy is enforced by the inversion symmetry. A coherent superposition of these modes forms a standing-wave configuration that cancels the net in-plane momentum, while preserving a uniform axial rotation. 

Our DFT calculations indeed reveal that two circularly polarized $E_g^{\pm}$ phonons with opposite polarizations along the $c$-axis remain degenerate (see Fig.~S5 in the Supplemental Material). Consequently, a condensate of these standing phonon waves yields a ferro-rotational order parameter $\mathcal{\bm{A}}$. Importantly, RIXS at momentum transfer $\bf{q}_\|=0$ selectively couples these standing waves. The non-reciprocal circular dichroism in RIXS manifests itself as a directional response upon reversal of X-ray wavevector $\bf{k}$.


\djh{The mechanism underlying the observed non-reciprocal CD in the phonon excitations is fundamentally distinct from that of magnetic and natural circular dichroism in X-ray absorption. In contrast, the non-reciprocal CD observed here occurs in the absence of both net magnetization and broken inversion symmetry. It originates from the coupling among photon helicity, phonon polarization, and the ferro-rotational axial order parameter. The dichroic signal emerges only in momentum-resolved RIXS, where this coupling gives rise to a non-reciprocal response.}

\begin{acknowledgments}
S.V.S. thanks T. Satoh for fruitful discussions. This work was supported in part by the National Science and Technology Council of Taiwan under Grant Nos. NSTC113-2112-M-007-033, NSTC113-2112-M-213-016, NSTC112-2112-M-007-031, and NSTC112-2112-M-213-026-MY3. This work was also supported by the W. M. Keck foundation grant to the Keck Center for Quantum Magnetism at Rutgers University. We also thank the support by the Japan Society for the Promotion of Science under Grant No. JP22K03535. Work at the University of Tennessee (curve fitting and writing) was supported by the U.S.~Department of Energy, Office of Science, Office of Basic Energy Sciences, under Award Number DE-SC0022311. DFT calculations were performed on the Uran supercomputer at the IMM UB RAS. SVS and EVK thank Ministry of Science and Higher Education of the Russian Federation for support, which came via Institute of Metal Physics. EVK was supported by the grant from the Foundation for the Development of Theoretical Physics and Mathematics "BASIS". MG acknowledges support from the Luxembourg National Research Fund under project LETZGROW (INTER/MOBILITY/2022/MS/17566170).
\end{acknowledgments}

\section*{Author Contributions}
D.J.H. conceived and led the project. H.Y.H., G.C., J.O., D.J.H., A.F., and C.T.C. conducted the RIXS measurements. M.G. and S.W.C. synthesized and characterized the samples. D.B. and S.J. performed the phonon model calculations. E.V.K. and S.V.S. conducted the DFT calculations. A.F. performed the symmetry analysis. C.Y.M. conducted the theoretical analysis. D.J.H., S.J., A.F., C.Y.M., and S.W.C. wrote the manuscript with input from the other authors.

\bibliography{ref}

@article{Gopalan2011rotation,
  title={Rotation-reversal symmetries in crystals and handed structures},
  author={Gopalan, Venkatraman and Litvin, Daniel B},
  journal={Nat. Mater.},
  volume={10},
  number={5},
  pages={376--381},
  year={2011},
  publisher={Nature Publishing Group UK London}
}

@article{Mitrano2024exploring,
  title = {Exploring Quantum Materials with Resonant Inelastic X-Ray Scattering},
  author = {Mitrano, M. and Johnston, S. and Kim, Young-June and Dean, M. P. M.},
  journal = {Phys. Rev. X},
  volume = {14},
  issue = {4},
  pages = {040501},
  numpages = {32},
  year = {2024},
  month = {Dec},
  publisher = {American Physical Society},
  doi = {10.1103/PhysRevX.14.040501},
  url = {https://link.aps.org/doi/10.1103/PhysRevX.14.040501}
}

@article{Geondzhian2020generalization,
  title = {Generalization of the {F}ranck-{C}ondon model for phonon excitations by resonant inelastic x-ray scattering},
  author = {Geondzhian, Andrey and Gilmore, Keith},
  journal = {Phys. Rev. B},
  volume = {101},
  issue = {21},
  pages = {214307},
  numpages = {12},
  year = {2020},
  month = {Jun},
  publisher = {American Physical Society},
  doi = {10.1103/PhysRevB.101.214307},
  url = {https://link.aps.org/doi/10.1103/PhysRevB.101.214307}
}

@article{Lee2013role,
  title = {Role of Lattice Coupling in Establishing Electronic and Magnetic Properties in Quasi-One-Dimensional Cuprates},
  author = {Lee, W. S. and Johnston, S. and Moritz, B. and Lee, J. and Yi, M. and Zhou, K. J. and Schmitt, T. and Patthey, L. and Strocov, V. and Kudo, K. and Koike, Y. and van den Brink, J. and Devereaux, T. P. and Shen, Z. X.},
  journal = {Phys. Rev. Lett.},
  volume = {110},
  issue = {26},
  pages = {265502},
  numpages = {5},
  year = {2013},
  month = {Jun},
  publisher = {American Physical Society},
  doi = {10.1103/PhysRevLett.110.265502},
  url = {https://link.aps.org/doi/10.1103/PhysRevLett.110.265502}
}

@article{thomas2024theory,
  title = {Theory of Electron-Phonon Interactions in Extended Correlated Systems Probed by Resonant Inelastic X-Ray Scattering},
  author = {Thomas, Jinu and Banerjee, Debshikha and Nocera, Alberto and Johnston, Steven},
  journal = {Phys. Rev. X},
  volume = {15},
  issue = {2},
  pages = {021030},
  numpages = {17},
  year = {2025},
  month = {Apr},
  publisher = {American Physical Society},
  doi = {10.1103/PhysRevX.15.021030},
  url = {https://link.aps.org/doi/10.1103/PhysRevX.15.021030}
}

@article{Johnson2012_PRL,
  title = {{Giant improper ferroelectricity in the ferroaxial magnet CaMn$_{7}$O$_{12}$}},
  author = {Johnson, R. D. and Chapon, L. C. and Khalyavin, D. D. and Manuel, P. and Radaelli, P. G. and Martin, C.},
  journal = {Phys. Rev. Lett.},
  volume = {108},
  issue = {6},
  pages = {067201},
  numpages = {4},
  year = {2012},
  month = {Feb},
  publisher = {American Physical Society},
  doi = {10.1103/PhysRevLett.108.067201},
  url = {https://link.aps.org/doi/10.1103/PhysRevLett.108.067201}
}

@article{Jin2020_NP,
  title={Observation of a ferro-rotational order coupled with second-order nonlinear optical fields},
  author={Jin, Wencan and Drueke, Elizabeth and Li, Siwen and Admasu, Alemayehu and Owen, Rachel and Day, Matthew and Sun, Kai and Cheong, Sang-Wook and Zhao, Liuyan},
  journal={Nat. Phys.},
  volume={16},
  number={1},
  pages={42--46},
  year={2020},
  publisher={Nature Publishing Group UK London}
}

@article{Cheong2018broken,
  title={Broken symmetries, non-reciprocity, and multiferroicity},
  author={Cheong, Sang-Wook and Talbayev, Diyar and Kiryukhin, Valery and Saxena, Avadh},
  journal={npj Quantum Mater.},
  volume={3},
  number={1},
  pages={19},
  year={2018},
  publisher={Nature Publishing Group UK London}
}

@article{Hayashida2020,
  title={Visualization of ferroaxial domains in an order-disorder type ferroaxial crystal},
  author={Hayashida, T and Uemura, Y and Kimura, K and Matsuoka, S and Morikawa, D and Hirose, S and Tsuda, K and Hasegawa, T and Kimura, T},
  journal={Nat. Commun.},
  volume={11},
  number={1},
  pages={4582},
  year={2020},
  publisher={Nature Publishing Group UK London}
}

@article{Ament11a,
	author = {Ament, L. J. P. and van Veenendaal, M. and van den Brink, J.},
	doi = {10.1209/0295-5075/95/27008},
	journal = {Europhys. Lett.},
	month = {jul},
	number = {2},
	pages = {27008},
	title = {Determining the electron-phonon coupling strength from Resonant Inelastic X-ray Scattering at transition metal {L}-edges},
	url = {https://dx.doi.org/10.1209/0295-5075/95/27008},
	volume = {95},
	year = {2011}
}

@article{Devereaux2016,
  title = {Directly Characterizing the Relative Strength and Momentum Dependence of Electron-Phonon Coupling Using Resonant Inelastic X-Ray Scattering},
  author = {Devereaux, T. P. and Shvaika, A. M. and Wu, K. and Wohlfeld, K. and Jia, C. J. and Wang, Y. and Moritz, B. and Chaix, L. and Lee, W.-S. and Shen, Z.-X. and Ghiringhelli, G. and Braicovich, L.},
  journal = {Phys. Rev. X},
  volume = {6},
  issue = {4},
  pages = {041019},
  numpages = {12},
  year = {2016},
  month = {Oct},
  publisher = {American Physical Society},
  doi = {10.1103/PhysRevX.6.041019},
  url = {https://link.aps.org/doi/10.1103/PhysRevX.6.041019}
}

@article{kidoh1984,
  title={{Electron density distribution in ilmenite-type crystals. II. Manganese (II) titanium (IV) trioxide}},
  author={Kidoh, KUMIKO and Tanaka, KIYOAKI and Marumo, FUMIYUKI and Takei, HUMIHIKO},
  journal={Acta. Crystallogr., Sect. B},
  volume={40},
  number={4},
  pages={329--332},
  year={1984},
  publisher={International Union of Crystallography}
}

@article{Sekine2024,
  title = {{Second harmonic imaging of antiferromagnetic domains and confirmation of absence of ferroaxial twins in MnTiO$_{3}$}},
  author = {Sekine, Daiki and Sato, Tatsuki and Tokunaga, Yusuke and Arima, Taka-hisa and Matsubara, Masakazu},
  journal = {Phys. Rev. Mater.},
  volume = {8},
  issue = {6},
  pages = {064406},
  numpages = {5},
  year = {2024},
  month = {Jun},
  publisher = {American Physical Society},
  doi = {10.1103/PhysRevMaterials.8.064406},
  url = {https://link.aps.org/doi/10.1103/PhysRevMaterials.8.064406}
}

@article{kishine2020,
  title={Chirality-induced phonon dispersion in a noncentrosymmetric micropolar crystal},
  author={Kishine, J. and Ovchinnikov, A. S. and Tereshchenko, A. A.},
  journal={Phys. Rev. Letts.},
  volume={125},
  number={24},
  pages={245302},
  year={2020},
  publisher={APS}
}

@article{YinAdvMater2021,
author = {Yin, Tingting and Ulman, Kanchan Ajit and Liu, Sheng and Granados del \'{A}guila, AndrÃes and Huang, Yuqing and Zhang, Lifa and Serra, Marco and Sedmidubsky, David and Sofer, Zdenek and Quek, Su Ying and Xiong, Qihua},
title = {Chiral Phonons and Giant Magneto-Optical Effect in {$\mathrm{CrBr}_3$} 2{$\mathrm{D}$} Magnet},
journal = {Adv. Mater.},
volume = {33},
number = {36},
pages = {2101618},
doi = {https://doi.org/10.1002/adma.202101618},
url = {https://onlinelibrary.wiley.com/doi/abs/10.1002/adma.202101618},
year = {2021}
}

@article{Zhu2018,
author = {Zhu, Hanyu and Yi, Jun and Li, Ming Yang and Xiao, Jun and Zhang, Lifa and Yang, Chih Wen and Kaindl, Robert A. and Li, Lain Jong and Wang, Yuan and Zhang, Xiang},
doi = {10.1126/science.aar2711},
issn = {10959203},
journal = {Science},
number = {6375},
pages = {579--582},
pmid = {29420291},
title = {{Observation of chiral phonons}},
volume = {359},
year = {2018}
}

@article{Ueda2023,
author = {Ueda, Hiroki and Garc{\'{i}}a-Fern{\'{a}}ndez, Mirian and Agrestini, Stefano and Romao, Carl P. and van den Brink, Jeroen and Spaldin, Nicola A. and Zhou, Ke Jin and Staub, Urs},
doi = {10.1038/s41586-023-06016-5},
isbn = {4158602306016},
issn = {14764687},
journal = {Nature},
number = {7967},
pages = {946--950},
pmid = {37286603},
publisher = {Springer US},
title = {{Chiral phonons in quartz probed by X-rays}},
volume = {618},
year = {2023}
}

@article{Ishito2023,
author = {Ishito, Kyosuke and Mao, Huiling and Kousaka, Yusuke and Togawa, Yoshihiko and Iwasaki, Satoshi and Zhang, Tiantian and Murakami, Shuichi and ichiro Kishine, Jun and Satoh, Takuya},
doi = {10.1038/s41567-022-01790-x},
issn = {17452481},
journal = {Nat. Phys.},
number = {1},
pages = {35--39},
publisher = {Springer US},
title = {{Truly chiral phonons in $\alpha$-HgS}},
volume = {19},
year = {2023}
}

@article{Lujan2024,
author = {David Lujan  and Jeongheon Choe  and Swati Chaudhary  and Gaihua Ye  and Cynthia Nnokwe  and Martin Rodriguez-Vega  and Jiaming He  and Frank Y. Gao  and T. Nathan Nunley  and Edoardo Baldini  and Jianshi Zhou  and Gregory A. Fiete  and Rui He  and Xiaoqin Li },
title = {Spin–orbit exciton–induced phonon chirality in a quantum magnet},
journal = {Proceedings of the National Academy of Sciences},
volume = {121},
number = {11},
pages = {e2304360121},
year = {2024},
doi = {10.1073/pnas.2304360121},
URL = {https://www.pnas.org/doi/abs/10.1073/pnas.2304360121},
}

@article{zhang2025,
  title = {{Directionally asymmetric nonlinear optics in ferrorotational MnTiO$_{3}$}},
  author = {Zhang, Xinshu and Carbin, Tyler and Du, Kai and Li, Bingqing and Wang, Kefeng and Li, Casey and Qian, Tiema and Ni, Ni and Cheong, Sang-Wook and Kogar, Anshul},
  journal = {Phys. Rev. B},
  volume = {112},
  issue = {12},
  pages = {L121108},
  numpages = {6},
  year = {2025},
  month = {Sep},
  publisher = {American Physical Society},
  doi = {10.1103/nsm8-6bm2},
  url = {https://link.aps.org/doi/10.1103/nsm8-6bm2}
}

@article{yokota2022,
  title={Three-dimensional imaging of ferroaxial domains using circularly polarized second harmonic generation microscopy},
  author={Yokota, Hiroko and Hayashida, Takeshi and Kitahara, Dan and Kimura, Tsuyoshi},
  journal={npj Quantum Mater.},
  volume={7},
  number={1},
  pages={106},
  year={2022},
  publisher={Nature Publishing Group UK London}
}

@article{Rossi2019experimental,
  title = {Experimental Determination of Momentum-Resolved Electron-Phonon Coupling},
  author = {Rossi, Matteo and Arpaia, Riccardo and Fumagalli, Roberto and Moretti Sala, Marco and Betto, Davide and Kummer, Kurt and De Luca, Gabriella M. and van den Brink, Jeroen and Salluzzo, Marco and Brookes, Nicholas B. and Braicovich, Lucio and Ghiringhelli, Giacomo},
  journal = {Phys. Rev. Lett.},
  volume = {123},
  issue = {2},
  pages = {027001},
  numpages = {6},
  year = {2019},
  month = {Jul},
  publisher = {American Physical Society},
  doi = {10.1103/PhysRevLett.123.027001},
  url = {https://link.aps.org/doi/10.1103/PhysRevLett.123.027001}
}

@article{ali2024,
  title={{Influence of anti-ferromagnetic ordering and electron correlation on the electronic structure of MnTiO$_3$}},
  author={Ali, Asif and Maurya, R. K. and Bansal, Sakshi and Reddy, B. H. and Singh, Ravi Shankar},
  journal={Europhys. Lett.},
  volume={147},
  number={4},
  pages={46002},
  year={2024},
  publisher={IOP Publishing}
}

@article{okamoto2025,
  title={Altermagnetic boosting of chiral phonons},
  author={Okamoto, J and Mou, C. Y. and Huang, H. Y. and Channagowdra, G. and Won, C. and Du, K. and Fang, X. and Komleva, E. V. and Chen, C. T. and Streltsov, S. V. and Fujimori, A. and Cheong, S.-W. and Huang, D. J.},
  journal={arXiv:2512.00388},
  year={2025}
}

@article{kusuno-prl,
  title = {{Raman optical activity induced by ferroaxial order in NiTiO$_{3}$}},
  author = {Kusuno, Gakuto and Hayashida, Takeshi and Nagai, Takayuki and Watanabe, Hikaru and Oiwa, Rikuto and Kimura, Tsuyoshi and Satoh, Takuya},
  journal = {Phys. Rev. Lett.},
  volume = {136},
  issue = {20},
  pages = {206902},
  numpages = {7},
  year = {2026},
  month = {May},
  publisher = {American Physical Society},
  doi = {10.1103/wrv8-4f7k},
  url = {https://link.aps.org/doi/10.1103/wrv8-4f7k}
}

@article{Furo2025PRB,
  title = {Theory of circular dichroism in resonant inelastic x-ray scattering},
  author = {Furo, M. and Hariki, A. and Kune\ifmmode \check{s}\else \v{s}\fi{}, J.},
  journal = {Phys. Rev. B},
  volume = {112},
  issue = {21},
  pages = {214429},
  numpages = {12},
  year = {2025},
  month = {Dec},
  publisher = {American Physical Society},
  doi = {10.1103/x96x-znr8},
  url = {https://link.aps.org/doi/10.1103/x96x-znr8}
}

@article{Goulon1998,
  title={{X-ray natural circular dichroism in a uniaxial gyrotropic single crystal of LiIO$_3$}},
  author={Goulon, Jos{\'e} and Goulon-Ginet, Chantal and Rogalev, Andrei and Gotte, Vincent and Malgrange, C{\'e}cile and Brouder, Christian and Natoli, Calogero R},
  journal={J. Chem. Phys.},
  volume={108},
  number={15},
  pages={6394--6403},
  year={1998},
  publisher={American Institute of Physics}
}

@article{Alagna1998,
  title={Natural circular dichroism},
  author={Alagna, L. and Prosperi, T. and Turchini, S. and Goulon, J. and Rogalev, A. and Goulon-Ginet, C. and Natoli, C. R. and Peacock, R. D. and Stewart, B.},
  journal={Phys. Rev. Lett.},
  volume={80},
  pages={4799},
  year={1998},
}

@article{Oreshko2018,
  title={X-ray natural circular dichroism in langasite crystal},
  author={Oreshko, Alexey P and Ovchinnikova, Elena N and Rogalev, Andrei and Wilhelm, Fabrice and Mill, Boris V and Dmitrienko, Vladimir E},
  journal={J. Synchrotron Radiat.},
  volume={25},
  number={1},
  pages={222--231},
  year={2018},
  publisher={International Union of Crystallography}
}

@article{Goulon2003,
  title={Optical activity probed with x-rays},
  author={Goulon, Jos{\'e} and Rogalev, Andrei and Wilhelm, Fabrice and Jaouen, Nicolas and Goulon-Ginet, Chantal and Brouder, Christian},
  journal={J. Phys.: Condens. Matter},
  volume={15},
  number={5},
  pages={S633},
  year={2003},
  publisher={IOP Publishing}
}

@article{okamoto2024giant,
  title={Giant X-Ray Circular Dichroism in a Time-Reversal Invariant Antiferromagnet},
  author={Okamoto, Jun and Wang, Ru-Pan and Chu, Yen-Yi and Shiu, Hung-Wei and Singh, Amol and Huang, Hsiao-Yu and Mou, Chung-Yu and Teh, Sukhito and Jeng, Horng-Tay and Du, Kai and Xu, Xianghan and Cheong, Sang-Wook and Du, Chao-Hung and Chen, Chien-Te and Fujimori, Atsushi  and Huang,  Di-Jing},
  journal={Adv. Mater.},
  volume={36},
  number={25},
  pages={2309172},
  year={2024},
  publisher={Wiley Online Library}
}

@article{huang2021quantum,
  title={Quantum fluctuations of charge order induce phonon softening in a superconducting cuprate},
  author={Huang, H. Y. and Singh, A. and Mou, C. Y. and Johnston, S. and Kemper, A. F. and van den Brink, J. and Chen, P. J. and Lee, T. K. and Okamoto, J. and Chu, Y. Y. and Li,  J. H. and Komiya, S. and Komarek, A. C. and Fujimori, A. and  Chen, C. T. and Huang, D. J.},
  journal={Phys. Rev. X},
  volume={11},
  number={4},
  pages={041038},
  year={2021},
  publisher={APS}
}

@article{Coh2023,
  title = {Classification of materials with phonon angular momentum and microscopic origin of angular momentum},
  author = {Coh, Sinisa},
  journal = {Phys. Rev. B},
  volume = {108},
  issue = {13},
  pages = {134307},
  numpages = {8},
  year = {2023},
  month = {Oct},
  publisher = {American Physical Society},
  doi = {10.1103/PhysRevB.108.134307},
  url = {https://link.aps.org/doi/10.1103/PhysRevB.108.134307}
}

@article{Zhang_chiral_phonon,
  title = {Chiral Phonons at High-Symmetry Points in Monolayer Hexagonal Lattices},
  author = {Zhang, Lifa and Niu, Qian},
  journal = {Phys. Rev. Lett.},
  volume = {115},
  issue = {11},
  pages = {115502},
  numpages = {5},
  year = {2015},
  month = {Sep},
  publisher = {American Physical Society},
  doi = {10.1103/PhysRevLett.115.115502},
  url = {https://link.aps.org/doi/10.1103/PhysRevLett.115.115502}
}

@article{Hayashida2021,
  title = {Phase transition and domain formation in ferroaxial crystals},
  author = {Hayashida, T. and Uemura, Y. and Kimura, K. and Matsuoka, S. and Hagihala, M. and Hirose, S. and Morioka, H. and Hasegawa, T. and Kimura, T.},
  journal = {Phys. Rev. Mater.},
  volume = {5},
  issue = {12},
  pages = {124409},
  numpages = {10},
  year = {2021},
  month = {Dec},
  publisher = {American Physical Society},
  doi = {10.1103/PhysRevMaterials.5.124409},
  url = {https://link.aps.org/doi/10.1103/PhysRevMaterials.5.124409}
}

@article{Mufti2011,
  title = {{Magnetoelectric coupling in MnTiO$_3$}},
  author = {Mufti, N. and Blake, G. R. and Mostovoy, M. and Riyadi, S. and Nugroho, A. A. and Palstra, T. T. M.},
  journal = {Phys. Rev. B},
  volume = {83},
  issue = {10},
  pages = {104416},
  numpages = {6},
  year = {2011},
  month = {Mar},
  publisher = {American Physical Society},
  doi = {10.1103/PhysRevB.83.104416},
  url = {https://link.aps.org/doi/10.1103/PhysRevB.83.104416}
}

@article{juraschek2025,
  title={Chiral phonons},
  author={Juraschek, Dominik M and Geilhufe, R Matthias and Zhu, Hanyu and Basini, Martina and Baum, Peter and Baydin, Andrey and Chaudhary, Swati and Fechner, Michael and Flebus, Benedetta and Grissonnanche, Gael and Kirilyuk, Andrei I.  and Lemeshko, Mikhail  and Maehrlein, Sebastian F. and Mignolet, Maxime and Murakami, Shuichi and Niu, Qian and Nowak, Ulrich and Romao, Carl P. and Rostami, Habib  and Satoh, Takuya and Spaldin, Nicola A. and Ueda, Hiroki and Zhang, Lifa},
  journal={Nat. Phys.},
   volume = {21},
  pages={1532–1540},
  year={2025},
  publisher={Nature Publishing Group UK London}
}

@article{He_2024,
  title = {Optical control of ferroaxial order},
  author = {He, Zhiren and Khalsa, Guru},
  journal = {Phys. Rev. Res.},
  volume = {6},
  issue = {4},
  pages = {043220},
  numpages = {12},
  year = {2024},
  month = {Dec},
  publisher = {American Physical Society},
  doi = {10.1103/PhysRevResearch.6.043220},
  url = {https://link.aps.org/doi/10.1103/PhysRevResearch.6.043220}
}

@article{Bhowal2024,
  title = {Electric toroidal dipole order and hidden spin polarization in ferroaxial materials},
  author = {Bhowal, Sayantika and Spaldin, Nicola A.},
  journal = {Phys. Rev. Res.},
  volume = {6},
  issue = {4},
  pages = {043141},
  numpages = {9},
  year = {2024},
  month = {Nov},
  publisher = {American Physical Society},
  doi = {10.1103/PhysRevResearch.6.043141},
  url = {https://link.aps.org/doi/10.1103/PhysRevResearch.6.043141}
}

@article{Martinez2025,
  title = {{Ferroaxial phonons in chiral and polar NiCo$_{2}$TeO$_{6}$}},
  author = {Martinez, V. A. and Gao, Y. and Yang, J. and Lyzwa, F. and Liu, Z. and Won, C. J. and Du, K. and Kiryukhin, V. and Cheong, S. W. and Sirenko, A. A.},
  journal = {Phys. Rev. B},
  volume = {112},
  issue = {6},
  pages = {064411},
  numpages = {6},
  year = {2025},
  month = {Aug},
  publisher = {American Physical Society}
}

@article{Wang2025alteraxial,
  title={Alteraxial Phonons in Collinear Magnets},
  author={Wang, Fuyi and Xu, Junqi and Liu, Xinqi and Wang, Huaiqiang and Zhang, Lifa and Zhang, Haijun},
  journal={arXiv:2512.07518},
  year={2025}
}

@article{Mustafa2025,
  title={{Origin of large effective phonon magnetic moments in monolayer MoS2}},
  author={Mustafa, Hussam and Nnokwe, Cynthia and Ye, Gaihua and Fang, Mengqi and Chaudhary, Swati and Yan, Jia-An and Wu, Kai and Cunningham, Connor J and Hemesath, Colin M and Stollenwerk, Andrew James and Shand, Paul M. and Yang, Eui-Hyeok and Fiete, Gregory A. and  He, Rui and Jin, Wencan},  
  journal={ACS nano},
  volume={19},
  number={11},
  pages={11241--11248},
  year={2025},
  publisher={ACS Publications}
}

@article{Tang2024,
  title = {{Exciton-activated effective phonon magnetic moment in monolayer ${\mathrm{MoS}}_{2}$}},
  author = {Tang, Chunli and Ye, Gaihua and Nnokwe, Cynthia and Fang, Mengqi and Xiang, Li and Mahjouri-Samani, Masoud and Smirnov, Dmitry and Yang, Eui-Hyeok and Wang, Tingting and Zhang, Lifa and He, Rui and Jin, Wencan},
  journal = {Phys. Rev. B},
  volume = {109},
  issue = {15},
  pages = {155426},
  numpages = {10},
  year = {2024},
  month = {Apr},
  publisher = {American Physical Society},
  doi = {10.1103/PhysRevB.109.155426},
  url = {https://link.aps.org/doi/10.1103/PhysRevB.109.155426}
}

\end{document}


\title{{\large Supplementary Information for} \\ Non-reciprocal circular dichroism of ferro-rotational phonons in MnTiO$_3$}

\author{H. Y. Huang, G. Channagowdra, D.~Banerjee, E.~V.~Komleva, J. Okamoto, C. T. Chen,\\ M. Guennou, S.~Johnston, S.~V.~Streltsov, C.~Y. Mou,  A. Fujimori,\\ S-W. Cheong\thanks{Corresponding author: sangc@physics.rutgers.edu}\, and D. J. Huang\thanks{Corresponding author: djhuang@nsrrc.org.tw}}

\date{\normalsize\today}

\maketitle
\hypersetup{%
    pdfborder = {0 0 0}
}
\vspace{3cm}
\begin{center}
{\Large \bf Contents}
\end{center}

\vspace{0.5em}

\noindent
\vspace{0.3em}

\noindent
\textbf{Table S1.} Irreducible representations of the \(C_{3i}\) point group with representative basis functions.

\vspace{0.3em}

\noindent
\textbf{Figure S1.} Orbital-selective phonon excitations by O K-edge RIXS.

\vspace{0.3em}

\noindent
\textbf{Figure S2.} Density of electronic states for MnTiO$_3$ calculated in DFT+U.

\noindent
\textbf{Figure S3.} Ferro-rotational phonons in MnTiO\(_3\) below \(T_{\mathrm{N}}\).

\vspace{0.3em}

\noindent
\textbf{Figure S4.} $Z$-component of the phonon circular polarization.

\vspace{0.3em}

\noindent
\textbf{Figure S5.} Crystal and magnetic structure of MnTiO$_3$.

\vspace{0.3em}

\noindent
\textbf{Figure S6.} Phonon density of states for MnTiO$_3$ calculated in DFT+U.

\vspace{0.3em}

\noindent
\textbf{Figure S7.} Phonon spectrum for MnTiO$_3$ calculated in DFT+U.

\vspace{0.3em}







\newpage
\setcounter{figure}{0}
\renewcommand{\thefigure}{S\arabic{figure}}

\setcounter{table}{0}
\renewcommand{\thetable}{S\arabic{table}}

\begin{table}[h!]
    \centering
    \begin{tabular}{cccccccc}
    \hline
     & $E$ & $I$ & $C_3$ & $C_3I$ & $C_3^2$ & $C_3^2I$ &Orbitals, operators  \\
    \hline
    $A_g$ & 1 & 1 & 1 & 1& 1 &1 & $z^2$, $x^2+y^2$, $L_z$  \\
    $E^+_g$ & 1 & 1 & $\omega^2$ & $\omega^2$& $-\omega$ &$-\omega$ &$x^2-y^2-ixy$, $L_x+iL_y$ \\
    $E^-_g$ & 1 & 1 & $-\omega$ &$-\omega$ & $\omega^2$ & $\omega^2$ &$x^2-y^2+ixy$, $L_x-iL_y$ \\
    $A_u$ & 1 & -1 &1 & -1& 1 &-1&$z$  \\
    $E^+_u$ & 1 & -1 &$\omega^2$ & $-\omega^2$& $-\omega$  &$\omega$    & $x+iy$  \\
    $E^-_u$ & 1 & -1 &$-\omega$  &$\omega$    & $\omega^2$ &$-\omega^2$ & $x-iy$\\
    \hline
    \end{tabular}
    \caption{Irreducible representation table of the $C_{3i}$ point group with representative basis functions. Note that $E^+_g$ and $E^-_g$ ($E^+_u$ and $E^-_u$) have distinct symmetries but energetically degenerate because they are mutually convertible via time reversal. $\omega\equiv e^{i\pi/3}$.}
    \label{irred-C3i}
\end{table}

\begin{figure}[h]
\centering
\includegraphics[width= 0.8 \columnwidth]{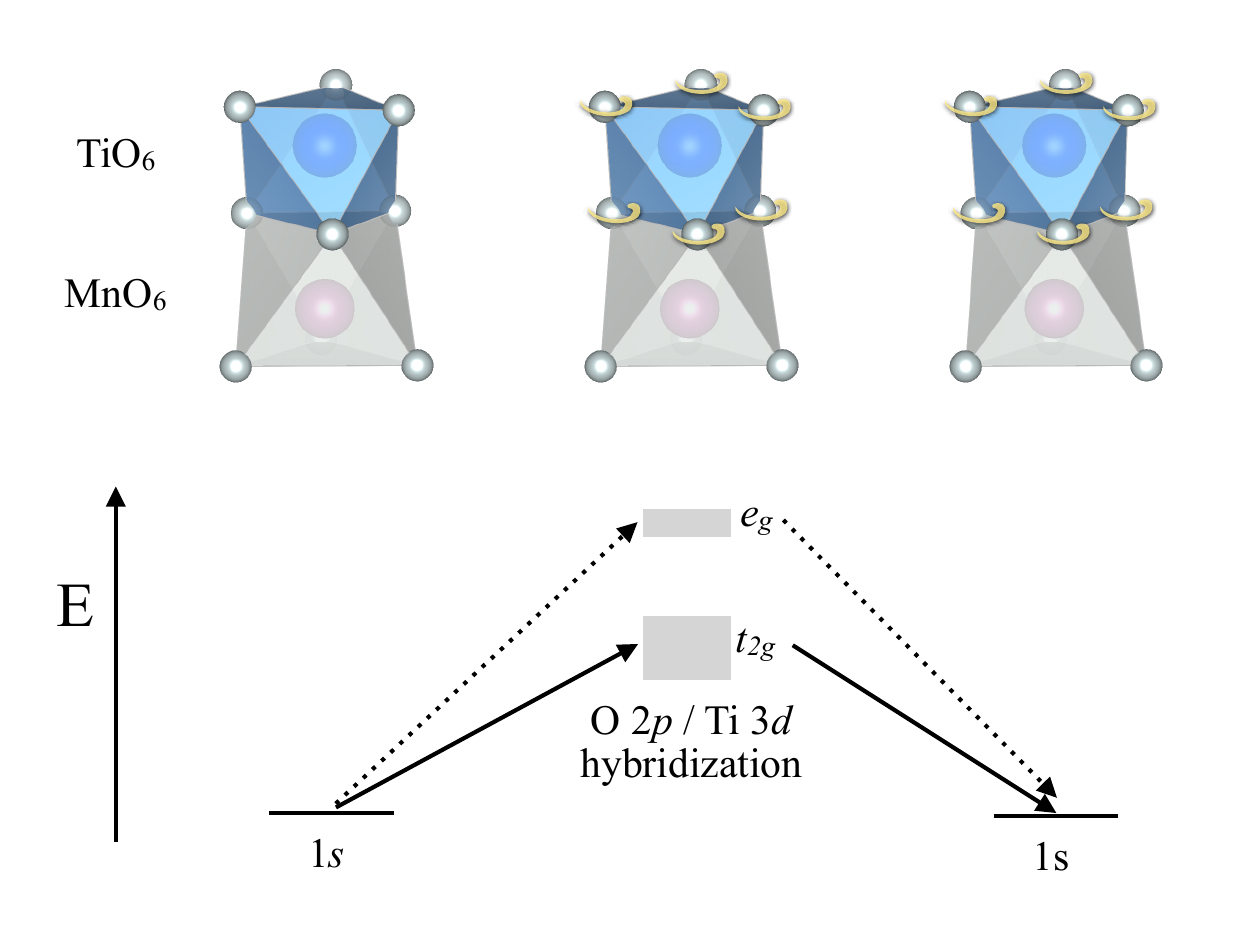}
\caption{{\bf Orbital-selective phonon excitations by O $K$-edge RIXS.} An incident photon excites an electron from the O 1$s$ core level into an unoccupied state with O 2$p$ character hybridized with Ti 3$d$  $t_{2g}$ or $e_g$ orbitals, forming a transient intermediate state. This state interacts with the lattice via electron-phonon coupling, leading to multiple phonon excitations. The excited electrons then recombine into the 1$s$ core hole, accompanied by the emission of a scattered photon. Tuning the incident photon energy to excite O $2p$ states hybridized with specific 3d orbitals enables orbital-selective phonon excitations.}\label{sf1}
\end{figure}

\begin{figure}[h]
\centering
\includegraphics[width=0.9\columnwidth]{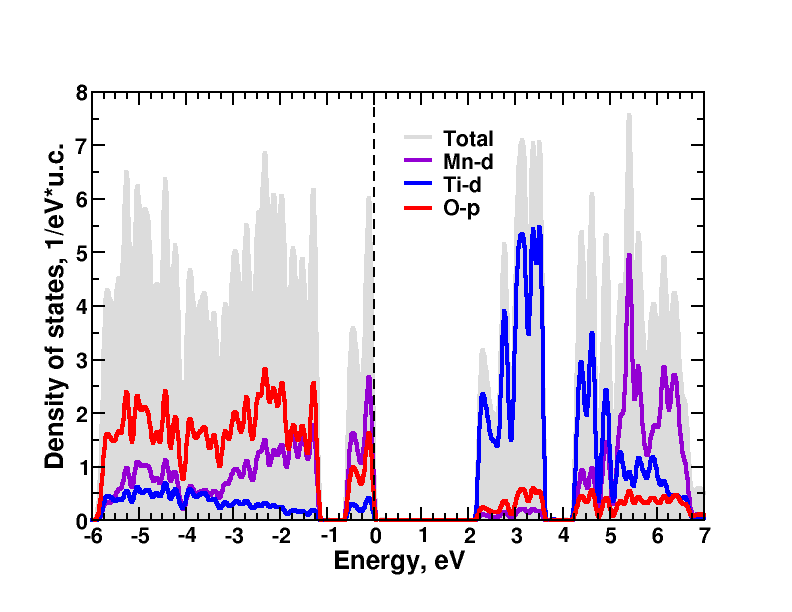}
\caption{{\bf Density of electronic states for MnTiO$_3$ calculated in DFT+U} for the cell containing 2 formula units. As antiferromagnetic order was considered, the results are given for one type of spin projection.}
\label{sf4}
\end{figure}

\begin{figure}[h]
\centering
\includegraphics[width=0.98\columnwidth]{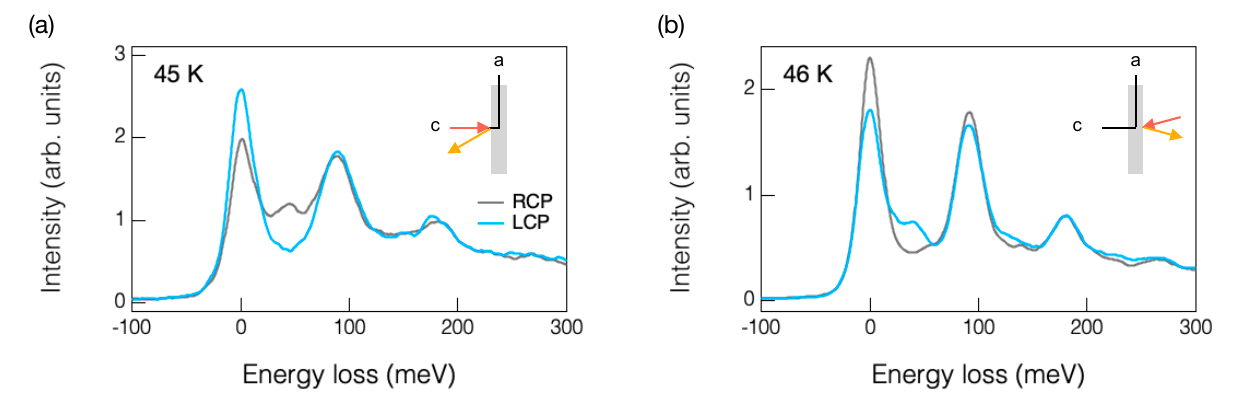}
\caption{{\bf Ferro-rotational phonons in MnTiO$_{3}$ below $T_{\text{N}}$.} RIXS spectra measured at different temperatures below \(T_{\mathrm{N}}\) using circularly polarized X-rays under various experimental conditions are shown. RCP and LCP denote right- and left-handed circularly polarized incident X-rays, respectively. The propagation directions of the incident and scattered X-rays (\(\omega_{\rm in}\) and \(\omega_{\rm out}\)) are indicated by red and orange arrows, respectively. The RIXS scattering angle, defined as the angle between the incident and scattered X-ray wavevectors, was fixed at \(150^\circ\). RIXS spectra were measured under normal-incidence conditions in (a). Spectra in (b) were collected with the X-ray beam incident on opposite faces of the sample with incident angle at \(75^\circ\).}
\label{sf2}
\end{figure}

\begin{figure}[h]
\centering
\includegraphics[width=0.9\columnwidth]{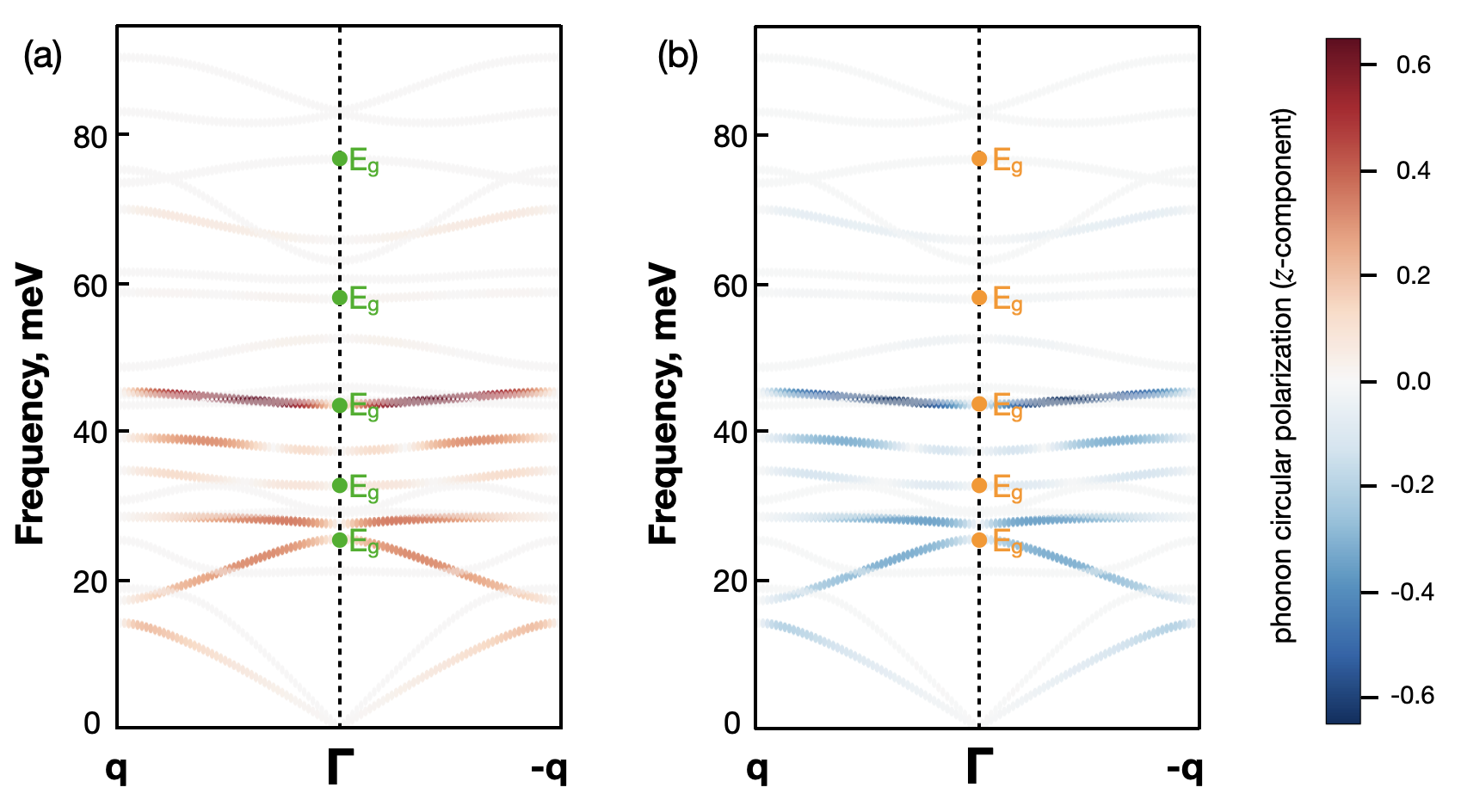}
\caption{{\bf $Z$-component of the phonon circular polarization} given for $E^+$ and $E^-$ representations of phonon modes, $\vec{q}=(\frac{1}{2},\frac{1}{2},\frac{1}{2})$ in fractional coordinates of a primitive cell.}
\label{sf7}
\end{figure}


\begin{figure}[h]
\centering
\includegraphics[width=0.9\columnwidth]{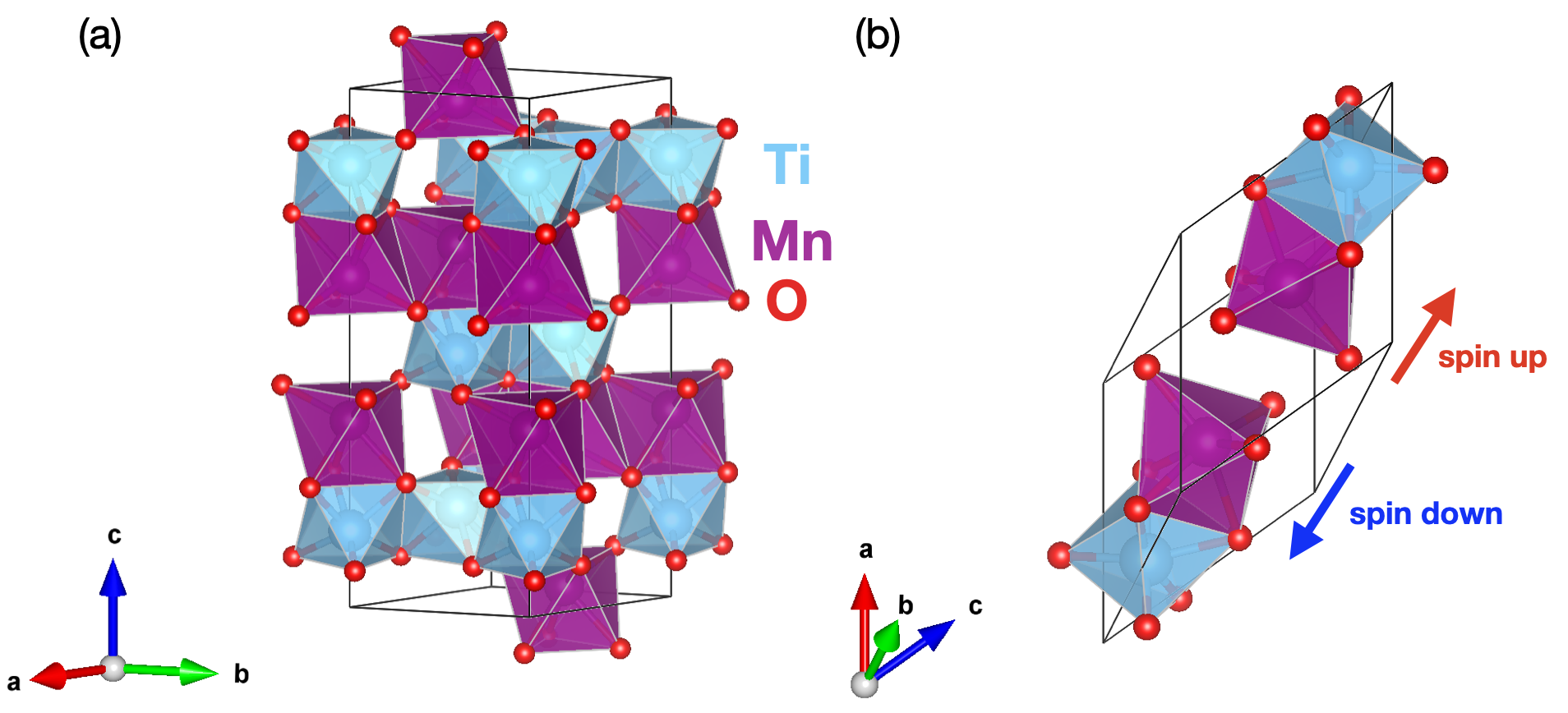}
\caption{{\bf Crystal and magnetic structure of MnTiO$_3$} with the corresponding (a) conventional and (b) primitive cells. The primitive one with 2 formula units was used for DFT+U calculations. Magnetic moments of Mn ions were arranged in the antiferromagnetic (all nearest neighbors antiferromagnetic) manner.}
\label{sf3}
\end{figure}

\begin{figure}[h]
\centering
\includegraphics[width=0.9\columnwidth]{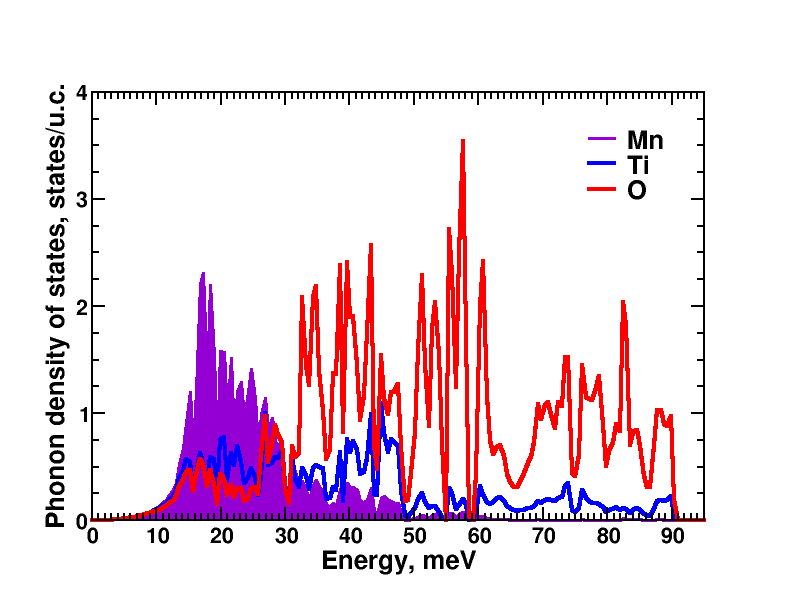}
\caption{{\bf Phonon density of states for MnTiO$_3$ calculated in DFT+U}.}
\label{sf5}
\end{figure}

\begin{figure}[h]
\centering
\includegraphics[width=0.9\columnwidth]{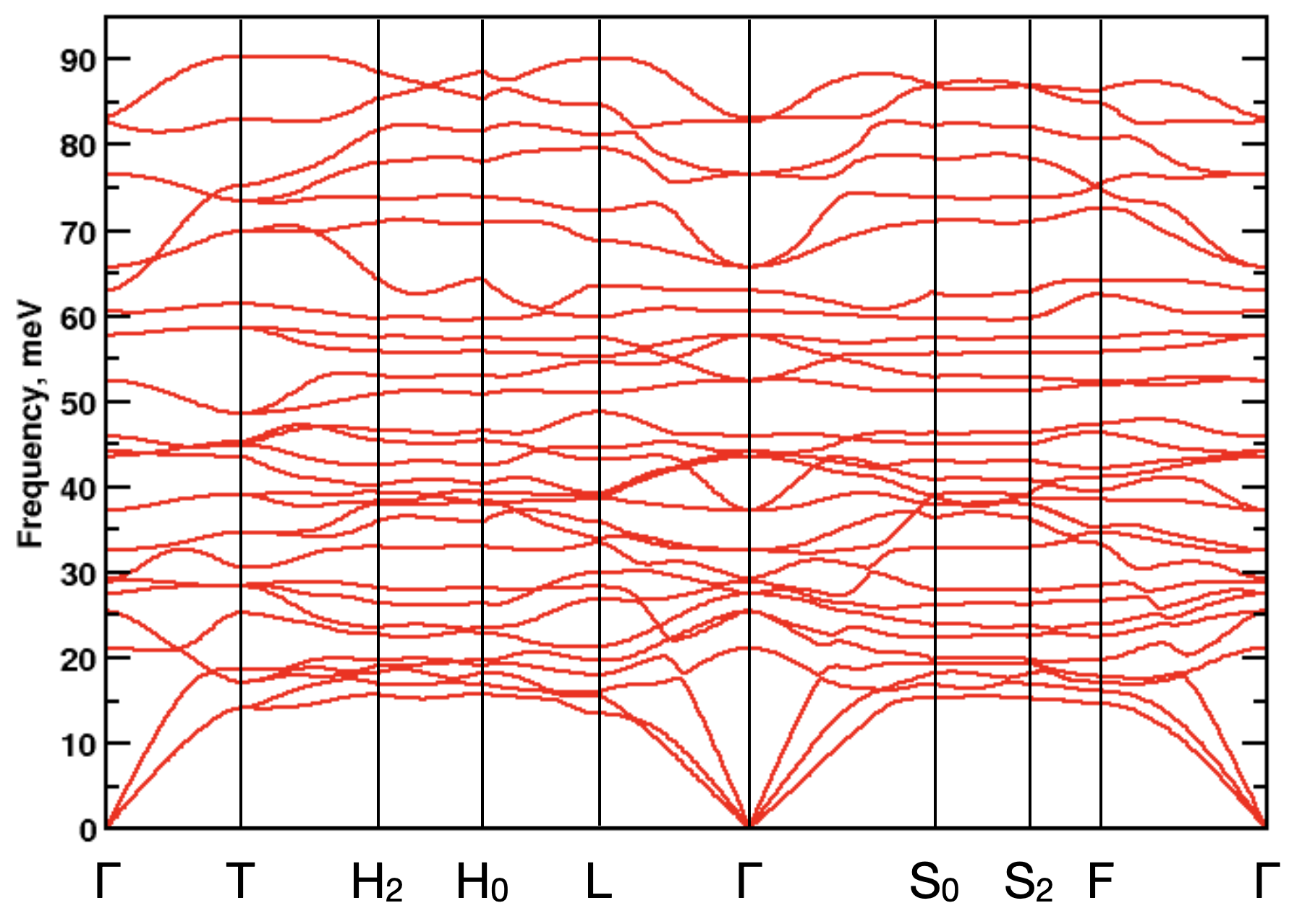}
\caption{{\bf Phonon spectrum for MnTiO$_3$ calculated in DFT+U}.}
\label{sf6}
\end{figure}